\documentclass[a4paper,twoside,10pt]{letter}
\usepackage{saj,graphicx,multicol,subeqnarray}
\usepackage{threeparttable}

\def\papertype{Original Scientific Paper}

\newcommand{\Halpha}{H${\alpha}$}

\def\p0{\phantom{0}}
\def\it{\sl}

\def\degr{\hbox{$^\circ$}}
\def\arcmin{\hbox{$^\prime$}}
\def\arcsec{\hbox{$^{\prime\prime}$}}
\def\SNR{\mbox{{SNR~J0519--6902}}}
\def\udc{524.354--77 : 524.722.3}
\begin{document}
\baselineskip=3.1truemm
\columnsep=.5truecm
\newenvironment{lefteqnarray}{\arraycolsep=0pt\begin{eqnarray}}
{\end{eqnarray}\protect\aftergroup\ignorespaces}
\newenvironment{lefteqnarray*}{\arraycolsep=0pt\begin{eqnarray*}}
{\end{eqnarray*}\protect\aftergroup\ignorespaces}
\newenvironment{leftsubeqnarray}{\arraycolsep=0pt\begin{subeqnarray}}
{\end{subeqnarray}\protect\aftergroup\ignorespaces}
%


\markboth{\eightrm Radio-continuum observations LMC SNR J0519--6902}
{\eightrm L.~M. Bozzetto et al. (2012) }

{\ }

\publ

\type

{\ }


\title{Radio-continuum observations of small, radially polarised Supernova Remnant J0519--6902 in the Large Magellanic Cloud} 


\authors{L.~M. Bozzetto,$^1$ M.~D.~Filipovi\'c,$^1$ D. Uro{\v s}evi{\'c}$^{2}$ \& E.~J.~Crawford$^1$}

\vskip3mm


%

\address{$^1$University of Western Sydney
 \break Locked Bag 1797, Penrith South DC, NSW 1797, Australia}  
\address{$^2$Department of Astronomy, Faculty of Mathematics, University of Belgrade,
 \break Studentski trg 16, 11000 Belgrade, Serbia}

\Email{m.filipovic@uws.edu.au}


\dates{October 2012}{November 2012}


\summary{We report on new Australian Telescope Compact Array (ATCA) observations of \SNR. The Supernova Remnant (SNR) is small in size ($\sim$8~pc) and exhibits a typical SNR spectrum of \mbox{$\alpha = -0.53\pm 0.07$}, with steeper spectral indices found towards the northern limb of the remnant. \SNR\ contains a low level of radially orientated polarisation at wavelengths of 3 \& 6~cm, which is characteristic of younger SNRs. A fairly strong magnetic field was estimated of $\sim$171~$\mu$G. The remnant appears to be the result of a typical Type~Ia supernovae, sharing many properties as another small and young Type~Ia LMC SNR, J0509--6731.}


\keywords{ISM: supernova remnants -- Magellanic Clouds -- Radio
  Continuum: ISM -- ISM: individual objects -- \SNR}

\begin{multicols}{2}
{


\section{1. INTRODUCTION}

Supernova Remnants (SNRs) play an essential role in the ecology of the universe, enriching the interstellar medium (ISM) as well as having a significant impact on the ISM structure and physical properties. The study of SNRs in our own galaxy isn't ideal due to the high level of absorption in addition to difficulties achieving accurate distance measurements. Instead, we look to the small dwarf galaxy, the Large Magellanic Cloud (LMC) for our study, which is located at a distance of 50~kpc (Macri et al. 2006). This proximity to us is far enough that we are able to assume all objects within the galaxy are located at the same distance, making measurements of extent \& surface brightness more reliable. There LMC also offers us an environment that is outside of the galactic plane at an angle of 35$^\circ$ (van der Marel \& Cioni~2001), and as a result, low foreground absorption.

One of the signatures of SNRs is their predominately non-thermal radio-continuum emission, typically exhibiting a spectrum of $\alpha\sim$ --0.5 (defined by $S\propto\nu^\alpha$). However, this value can vary as there is a wide variety of SNRs in different stages of evolution (Filipovi\'c et al. 1998; Payne et al. 2008)

\SNR\ was observed by Tuohy et al. (1982) who recorded an integrated flux density measurement at 408~MHz of 150$\pm$30~mJy and 33$\pm$5~mJy at 5000~MHz. They make note that the SNR is Balmer dominated with a broad \Halpha\ component and reported a X-ray extent of $\sim$30\arcsec\ and an optical extent of 28\arcsec. They estimate a shock velocity of 2900$\pm$400 km s$^{-1}$ and an age of $\sim$500 years. It is also mentioned that the remnant is expanding into a low density region composed of neutral hydrogen; inferring Type~I supernovae and they estimate the progenitor mass to be between 1.2 and 4 solar masses. Mathewson et al. (1983) recorded a spectral index of --0.6. Mills et al. 1984 record a 843~MHz flux density measurement of 145~mJy, updating the spectrum of the remnant to --0.65. Chu \& Kennicutt 1988 associated this SNR with the nearby (200~pc) OB -- LH41 and classified it as population~II. Smith et al. (1991)  estimated an age between 500 -- 1500 years, noting that this makes remnant one of the youngest in the LMC. Dickel \& Milne (1994) state that this SNR is similar to Tycho or Keplers SNR in the Milky Way. Dickel \& Milne (1995) then observed this remnant at wavelengths of 20 \& 13~cm, obtaining integrated flux densities of 100~mJy \& 70~mJy respectively. They also estimate mean fractional polarisation across the remnant of 1.5$\pm$0.6\% (20~cm) and 4.1$\pm$0.6\% (13~cm), and make note of the radial magnetic field and similarities to other young galactic SNRs. Filipovi\'c et al. (1995) measured an integrated flux density measurement of 57~mJy at 3~cm. Filipovi\'c et al. (1998) reobserved this SNR with the Parkes radio-telescope at 6~cm, estimating an integrated flux density of 72~mJy. Haberl \& Piestch (1999) observed this SNR with the ROSAT and gave the association [HP] 789. Borkowski et al. (2006) place this SNR at 600 years old with a 30\% error. Vukoti\'c et al. (2007) estimated the magnetic field of this SNR using the classical equipartition formula (186 $\mu$G) as well as a revised equipartition formula (270 $\mu$G). Desai et al. 2010 found no young stellar object (YSO) associated with this SNR. Kosenko et al (2010) used Chandra and the XMM Newton telescopes to estimate an age of 450$\pm$200 for this remnant. Most recently, Edwards et al. (2012) state that based on their current models, this SNR could have only been the result of a supersoft source or a double degenerate system mimicking a Type~I SN event.

In this paper we present new radio-continuum measurements and polarmetric analysis of \SNR. The observations, data reduction and imaging techniques are described in Section~2. The astrophysical interpretation of newly obtained moderate-resolution total intensity images are discussed in Section~3.

\section{2. OBSERVATIONS}

We observed \SNR\ on the 15$^\mathrm{th}$ and 16$^\mathrm{th}$ of November 2011 with the Australian Telescope Compact Array (ATCA), using the new Compact Array Broadband Backend (CABB) receiver at array configuration EW367, at wavelengths of 3 and 6~cm ($\nu$=9000 and 5500~MHz). Baselines formed with the $6^\mathrm{th}$ ATCA antenna were omitted, as the other five antennas were arranged in a compact configuration. The observations were carried out in the so called ``snap-shot'' mode, totaling $\sim$50 minutes of integration over a 14 hour period. Source PKS~B1934-638 was used for primary calibration and source PKS~B0530-727 was used for secondary (phase) calibration. The \textsc{miriad} (Sault et al.~1995) and \textsc{karma} (Gooch~1995) software packages were used for reduction and analysis. More information on the observing procedure and other sources observed in this session/project can be found in Boji\v{c}i\'c~et~al.~(2007), Crawford~et~al.~(2008a,b; 2010), \v{C}ajko~et~al.~(2009), De Horta et al. (2012), Grondin et al. (2012), Maggi et al. (2012) and Bozzetto et al.~(2010; 2012a,b,c,d).

Images were formed using \textsc{miriad} multi-frequency synthesis (Sault and Wieringa~1994) and natural weighting. They were deconvolved using the {\sc mfclean} and {\sc restor} algorithms with primary beam correction applied using the {\sc linmos} task. A similar procedure was used for both \textit{U} and \textit{Q} Stokes parameter maps. 

In addition to our own observations, we made use of two ATCA projects (C354 \& C149) at wavelengths of 13 \& 20~cm. Observations from project C354 were taken on the 18$^\mathrm{th}$ (array 1.5B), 22$^\mathrm{nd}$ and 23$^\mathrm{rd}$ (array 1.5D) of September 1994. Observations from project C149 were taken on the 22$^\mathrm{nd}$ of March (array 6A) and the 2$^\mathrm{nd}$ of April (array 6C).

\section{3. RESULTS AND DISCUSSION}

\SNR\ exhibits a ring-like shell morphology, with three brightened regions towards the northern, southern and eastern limb of the remnant (Fig.~1). The SNR is centred at RA(J2000)=5$^h$19$^m$34.9$^s$, DEC(J2000)=--69\degr02\arcmin07.9\arcsec. We estimate the spatial extent of \SNR\ at the 3$\sigma$ (Table~1; Col.~2) level (0.9~mJy) along the major (N-S) and minor (E-W) axes. Its size at 13~cm is 34\arcsec$\times$34\arcsec$\pm$4\arcsec\ (8$\times$8~pc with 1~pc uncertainty in each direction). We also estimate the ring thickness of the remnant to $\sim$7.3\arcsec\ (1.8~pc) at 13~cm, about 43\% of the SNR's radius. 


We use the integrated flux density measurements in Table~1 to estimate the radio spectral index of this remnant ($\alpha$=$-$0.53$\pm$0.07). The 408~MHz value has an error of 20\%, while we assume a 10\% error for the remaining values. The spectrum of this remnant is steeper than that as measured by Dickel \& Milne 1995 of --0.44. However, they make note that there was uncertainty in this measurement. To see the change in flux across the remnant, we created a spectral map image (Fig.~3) between 20 \& 13~cm wavelengths. The map was produced by reprocessing both 20 \& 13~cm images to a common {\it u -- v} range and then fitting $S\propto\nu^\alpha$ pixel by pixel in both images simultaneously. The emission falls predominately between --0.5 and --0.8, which is what we would generally expect of a younger SNR and is consistent with the overall spectrum of this image (--0.53).

\vspace{3mm}

\centerline{{\bf Table 1.} Flux Density of SNR J0519-6902.}
\vskip2mm
\centerline{
\begin{tabular}{ccccl}
\hline
$\lambda$ & R.M.S  & Beam Size  & S$_\mathrm{Total}$\\
(cm)      & (mJy) & (\arcsec) & (mJy)\\
\hline
73 & -- & 157$\times$172 & 150.0\\
36 & 0.5 & 46$\times$43 & 145.0\\
20 & 0.3 & 20$\times$19 & 121.8\\
13 & 0.3 & 6$\times$5 & 78.5\\
6 & 0.3 & 38$\times$24 & 46.5\\
3 & 0.3 & 23$\times$16 & 33.0\\
\hline
\end{tabular}}
\vspace{0.5cm}

Linear polarisation images were created at 6 \& 3~cm using \textit{Q} and \textit{U} stoke parameters (Fig.~4 \& 5). The mean fractional polarisation was calculated using flux density and polarisation:\\
\vspace{3mm}

\centerline{P=$\frac{\sqrt{S_{Q}^{2}+S_{U}^{2}}}{S_{I}}\cdot 100\%$}
\vspace{3mm}
\noindent where $S_{Q}, S_{U}$ and $S_{I}$ are integrated intensities for the \textit{Q}, \textit{U} and \textit{I} Stokes parameters. Our estimated peak value is 8.1\%$\pm$2.2\% (7$\sigma$) at 6~cm and 9.3\%$\pm$4.8\% (6$\sigma$) at 3~cm. The polarisation from the remnant appears radial at both wavelength. We estimate a mean polarisation across the remnant of $\sim$2.2\% at 6~cm and $\sim$3.2\% at 3~cm.

This radial polarisation is expected of smaller, younger SNRs and is comparable to similarly small SNRs in our own galaxy (Tycho's SNR -- Dickel et al. 1991) and SNRs in the LMC (SNR 0509-6731 -- Bozzetto et al. in prep). The remnants low level of radial polarisation is consistent with previous polarisation studies of \SNR\ at wavelengths of 20 \& 13~cm by Dickel et al. (1995), who also found this radially orientated polarisation, with a mean fractional polarisation of 1.5$\pm$0.6\% (20 cm) and 4.1$\pm$0.6\% (13 cm) across the remnant.

The polarisation position angles from these 6 \& 3~cm observations were used to estimate the Faraday rotation across the remnant (Fig.~6). Filled squares represent positive rotation measure, and open boxes negative rotation measure. Average rotation measure across the entire remnant was estimated at $\sim$10~rad~m$^{-2}$. However, as there was a significant amount of both positive and negative rotation measure across the remnant, we have broken down the remnant into three regions and performed separate analysis on each. Field 01 (North-East region) -- This region is dominated by negative rotation measure with an average value of --272~rad~m$^{-2}$, and a peak of --431~rad~m$^{-2}$. Field 02 (North-West region) -- In contrast, this region is dominated by positive rotation measure, with an average value of 462~rad~m$^{-2}$ and a peak of 624~rad~m$^{-2}$. Field 03 (Southern region) -- Similarly, this region is also dominated by positive rotation measure, with an average value of 697~rad~m$^{-2}$ and a peak of 784~rad~m$^{-2}$. The two southern most rotation measure pixels in this image were omitted from analysis as they were towards the edge of the remnant, where the polarised intensity is too weak to measure accurate rotation measure.

We used the new equipartition formula for SNRs (Arbutina et al. 2012), to estimate the magnetic field strength for the \SNR. The derivation of the new equipartition formula is based on the Bell (1978) diffuse shock acceleration (DSA) theory. This derivation is purely analytical, accommodated especially for the estimation of magnetic field strength in SNRs. The average equipartition field over the whole shell of \SNR\ is $\sim$171~$\mu$G with an estimate of E$_{min}$ = 1.82$\times$10$^{49}$~ergs (see Arbutina et al. (2012); and corresponding "calculator"\footnote{The calculator is available on http://poincare.matf.bg.ac.rs/\~{}arbo/eqp/ }). This value is typical for young SNRs with a strongly amplified magnetic field.

The surface brightness--diameter  ($\Sigma-D$) relationship for this SNR can be seen in Fig.~7 at 1~GHz with theoretically-derived evolutionary tracks (Berezhko \& V\"olk) superposed. \SNR\ is positioned at $(\Sigma, D)$ = ($5.5\times~10^{-20}$~W~m$^{-2}$~Hz$^{-1}$~Sr$^{-1}$, 8.2~pc) on the diagram. The location on the diagram shows that it is young SNR, in the early Sedov phase of evolution. Also, this object evolves in a rare environment and initial energy of explosion was low.

\vskip5mm

}

\end{multicols}


\centerline{\includegraphics[trim=0 0 0 0,angle=-90,width=.6\textwidth]{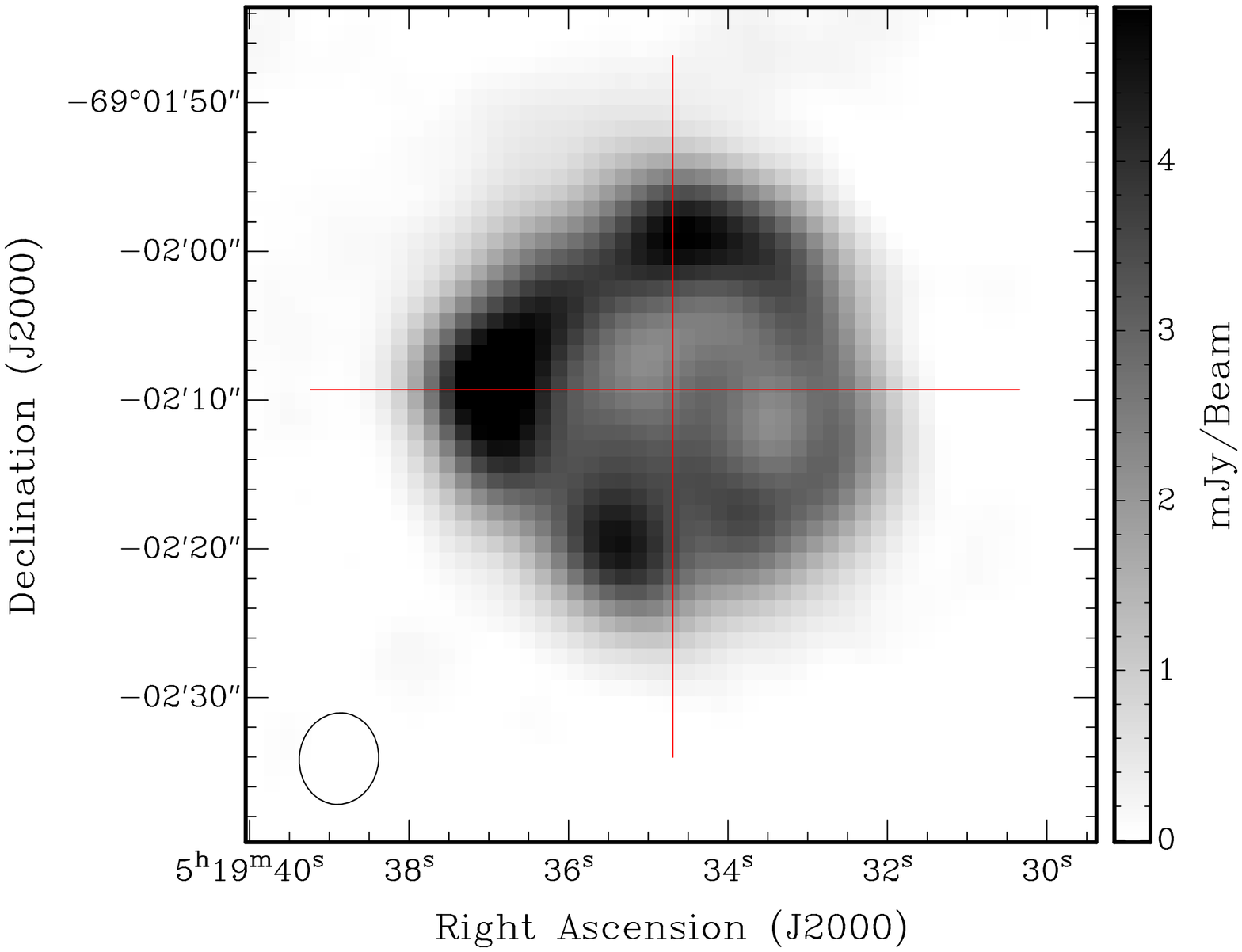}}
\centerline{\includegraphics[trim=0 0 0 0,width=.45\textwidth]{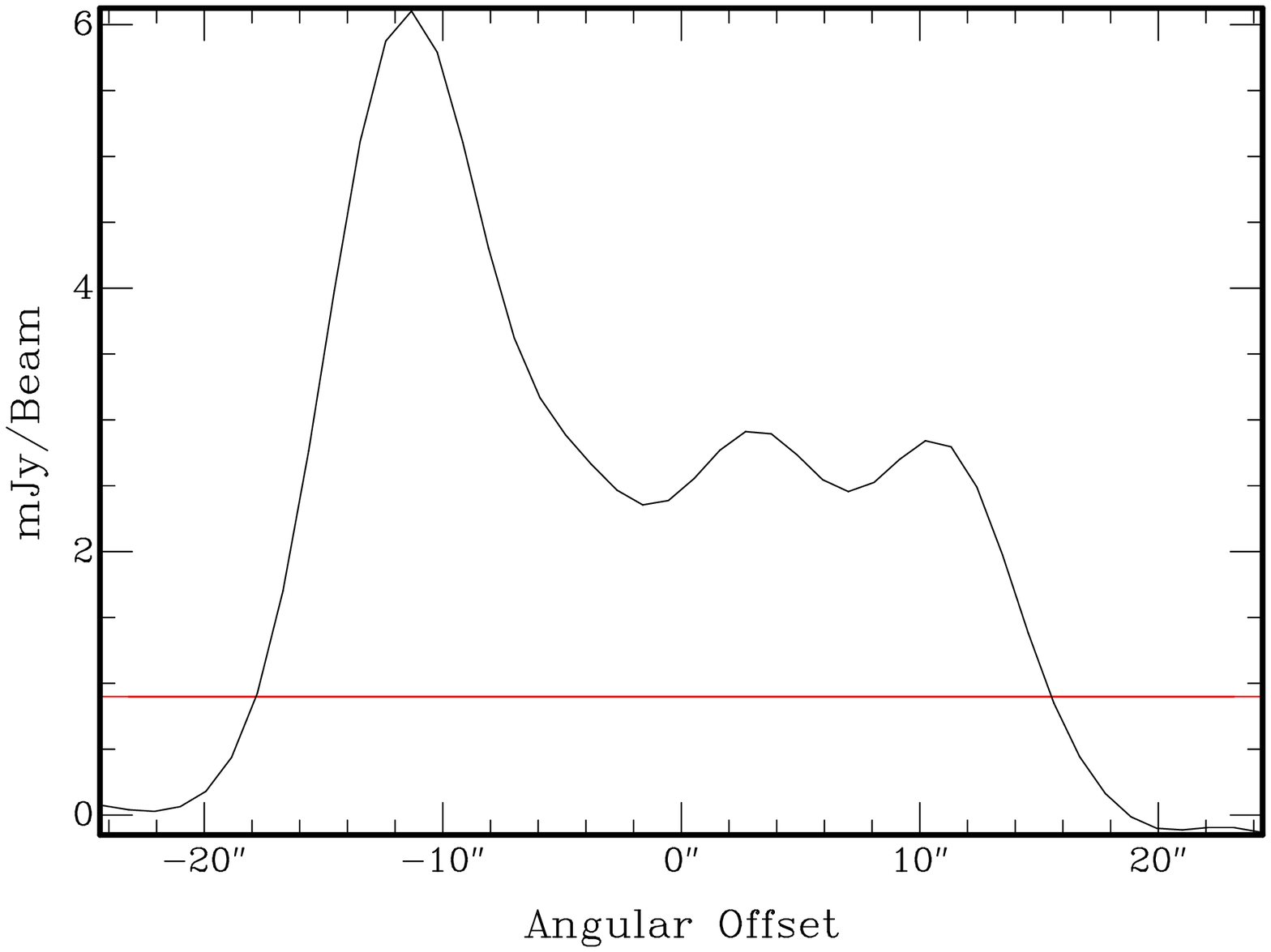}}
\centerline{\includegraphics[trim=0 0 0 0,width=.45\textwidth]{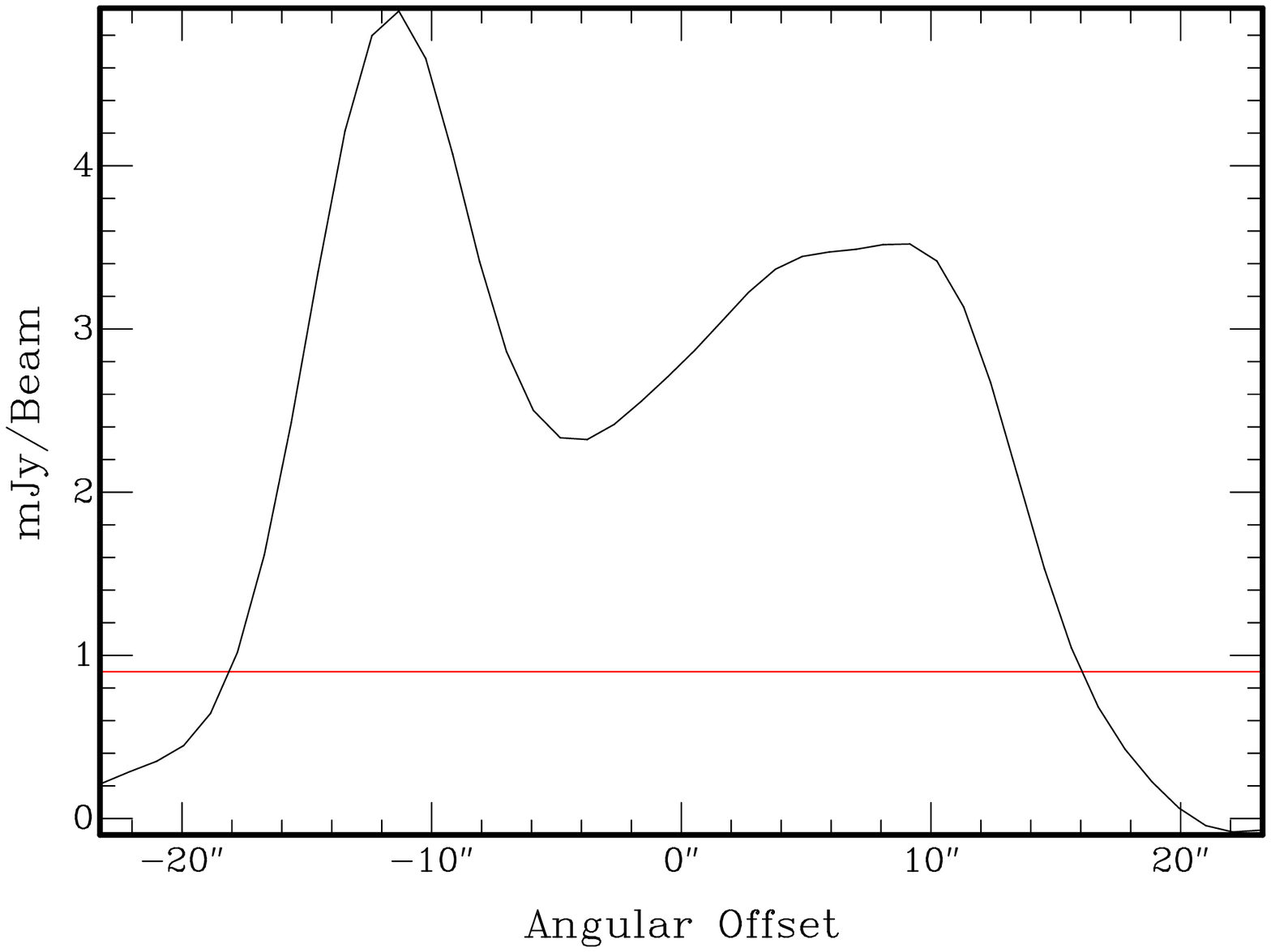}}
\figurecaption{1.}{The top image is an ATCA image of \SNR\ overlaid with major (EW) and minor (NS) axis. The middle and lower image show the flux emission at the major and minor axis respectively, with an overlaid line at 3$\sigma$.}
%

\centerline{\includegraphics[angle=-90, scale=.5]{0519-spcidx}}
\figurecaption{2.}{Radio-continuum spectrum of \SNR.}
\label{spcidx}

 \centerline{\includegraphics[angle=-90, scale=.5, trim=0 0 0 30, clip]{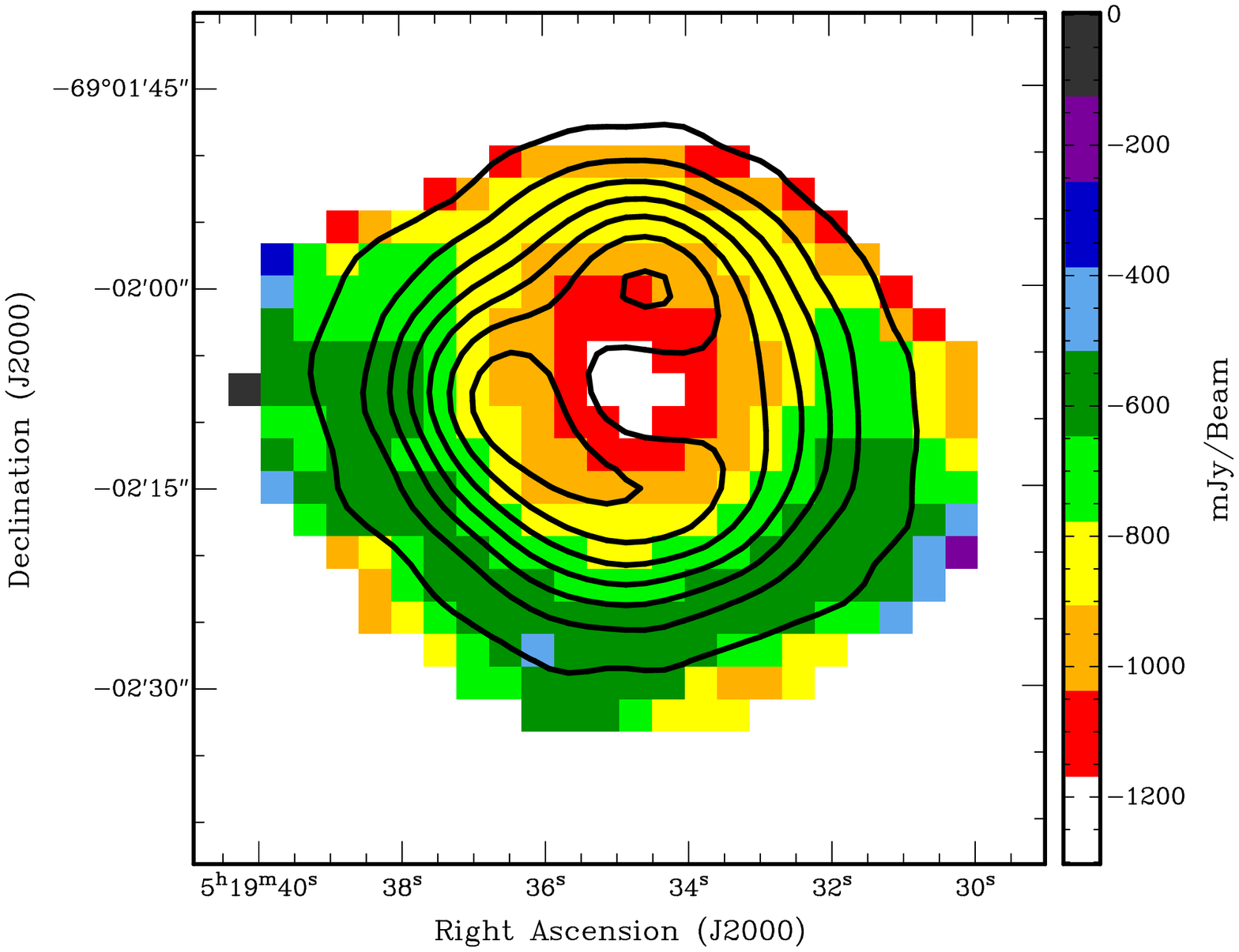}}
\figurecaption{3.}{Spectral map of \SNR\ between 20 \& 13~cm with overlaid contours at 13~cm of 3, 9, 15, 21, 27, 33 \& 39$\sigma$. The sidebar quantifies the change in spectral index. For example: -200 represents $\alpha$=--0.2.}
 \label{spcidx}

\centerline{\includegraphics[angle=-90, scale=.45, trim=0 0 0 0, clip]{0519-6902-6cm-pol}}
\figurecaption{4.}{Polarisation vectors overlaid on 6~cm ATCA observations of \SNR. The contours used are 3, 23, 43, 63, 83 \& 103$\sigma$. The blue ellipse in the lower left corner represents the synthesised beamwidth of 23.1\arcsec$\times$16.1\arcsec and the blue line below the ellipse represents a polarisation vector of 100\%.}
\label{polar}
 
 \centerline{\includegraphics[angle=-90, scale=.45, trim=0 0 0 0, clip]{0519-6902-3cm-pol}}
\figurecaption{5.}{Polarisation vectors overlaid on 3~cm ATCA observations of \SNR. The contours used are 3, 23 \& 43$\sigma$. The blue ellipse in the lower left corner represents the synthesised beamwidth of 38.0\arcsec$\times$24.6\arcsec and the blue line below the ellipse represents a polarisation vector of 100\%.}
\label{polar}

\centerline{\includegraphics[angle=-90, scale=.38, trim=0 0 0 0, clip]{0519-6902-rm}}
\figurecaption{6.}{Faraday rotation measure of \SNR\ overlaid on 3~cm image contours. The contours used are 3, 23 \& 43$\sigma$. Filled squares represent positive rotation measure while open squares represent negative rotation measure. The ellipse in the lower left represents the synthesised beamwidth of 38.0\arcsec$\times$24.6\arcsec. }
\label{polar}

\centerline{\includegraphics[scale=.9, trim=150 250 200 250, clip]{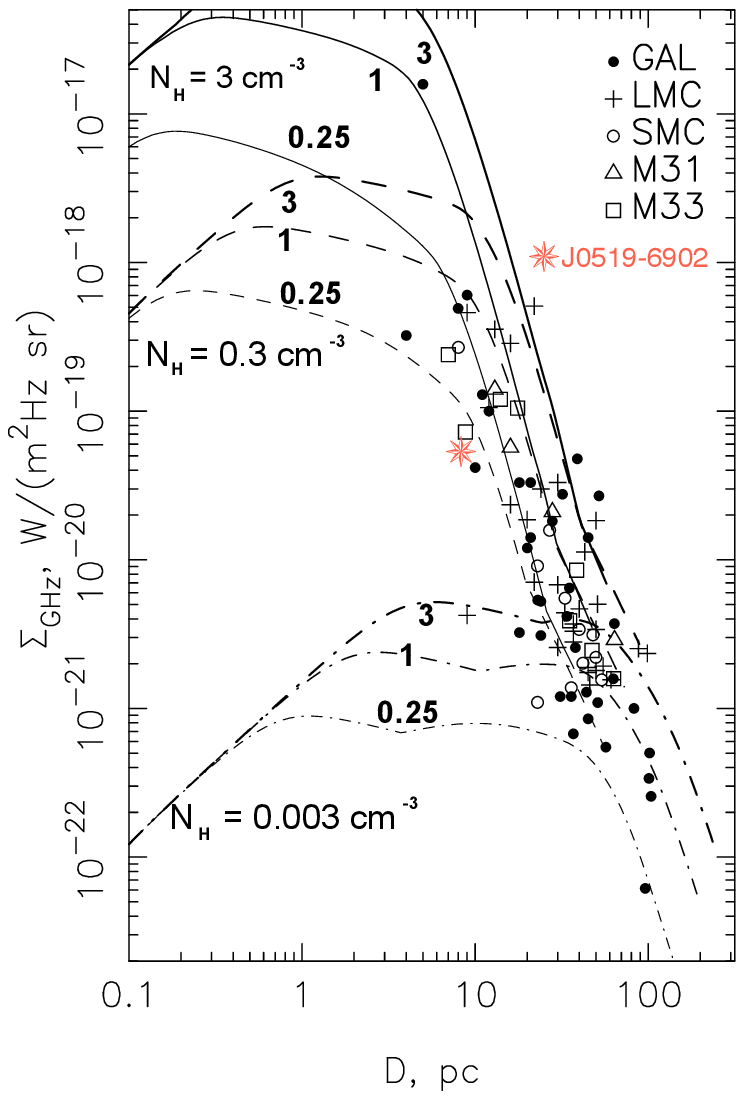}}
\figurecaption{7.}{Surface brightness-to-diameter diagram from Berezhko \& V\"olk (2004), with \SNR\ added. The evolutionary tracks are for ISM densities of N$_{H}$= 3, 0.3 and 0.003~cm$^{-3}$ and explosion energies of E$_{SN}$ = 0.25, 1 and 3$\times10^{51}$~erg.}
\label{polar}

%
%
%

\begin{multicols}{2}

{

\section{4. CONCLUSION}

\vskip-1mm

This remnant appears to exhibit a ring-like shell morphology with an extent of D=(8$\times$8)$\pm$1~pc, radially polarisation at 6 \& 3~cm with a mean integrated polarisation of $\sim$2.2\% \& $\sim$3.2\% respectively, a typical spectrum of $\alpha$ = --0.53$\pm$0.07, areas of positive (mean = 544~rad~m$^{-2}$) and negative \mbox{(mean = --290~rad~m$^{-2}$)} rotation measure. The estimated value of the magnetic field and location on the $\Sigma-D$ diagram show that this SNR is young, in the early Sedov phase of evolution. It expands in a less dense environment and the initial energy of probably Type~Ia explosion was low.


\acknowledgements{We used the {\sc karma} software package developed by the ATNF. The Australia Telescope Compact Array is part of the Australia Telescope which is funded by the Commonwealth of Australia for operation as a National Facility managed by CSIRO. This research is supported by the Ministry of Education and Science of the Republic of Serbia through project No. 176005}

\vskip.6cm


\references

\newcommand{\MNRAS}{\journal{Mon. Not. R. Astron. Soc.}}
\newcommand{\ApJ}{\journal{Astrophys. J.}}
\newcommand{\ApJS}{\journal{Astrophys. J. Supplement}}
\newcommand{\AJ}{\journal{Astronomical. J.}}

\vskip-1mm

{Arbutina}, B., {Uro{\v s}evi{\'c}}, D., {Andjeli{\'c}}, M.~M., {Pavlovi{\'c}}, M.~Z. and {Vukoti{\'c}}, B., 2012, \ApJ, \vol{746}, 79
%

Bell A. R., 1978, \MNRAS, \vol{182}, 443

{Berezhko}, E.~G. and {V{\"o}lk}, H.~J., 2004, \journal{Astron. Astrophys.}, \vol{427}, 525

%
Boji{\v c}i{\'c}, I.~S., Filipovi{\'c}, M.~D., Parker, Q.~A., Payne, J.~L., Jones, P.~A., Reid, W., Kawamura, A., Fukui, Y.: 2007, \MNRAS, \vol{378}, 1237.

{Borkowski}, K.~J., {Hendrick}, S.~P. and {Reynolds}, S.~P., 2006, \ApJ, 652, 1259

Bozzetto, L.~M., Filipovi{\'c}, M.~D., Crawford, E.~J., Boji{\v c}i{\'c}, I.~S., Payne, J.~L., Mendik, A., Wardlaw, B. and de Horta, A.~Y.,\ 2010, \journal{Serb. Astron. J.}, \vol{181}, 43

Bozzetto, L.~M., Filipovi{\'c}, M.~D., Crawford, E.~J., Payne, J.~L., De Horta, A.~Y. and Stupar, M.,\ 2012, RMxAA, \vol{48}, 41

Bozzetto, L.~M., Filipovi{\'c}, M.~D., Crawford, E.~J., Haberl, F., Sasaki, M., Uro{\v s}evi{\'c}, D., Pietsch, W., Payne, J.~L., de Horta, A.~Y., Stupar, M., Tothill, N.~F.~H., Dickel, J., Chu, Y.-H. and Gruendl, R.,\ 2012, MNRAS, 420, 2588

{Bozzetto}, L.~M., {Filipovic}, M.~D., {Crawford}, E.~J., {De Horta}, A.~Y. and {Stupar}, M., 2012, Serbian Astronomical Journal, \vol{184}, 69

\v{C}ajko K.~O., Crawford E.~J., Filipovi{\'c}, M.~D.: 2009, \journal{Serb. Astron. J.}, \vol{179}, 55. 

Chu, Y-H., and Kennicutt, R.~C.: 1988, \AJ, \vol{96}, 1874.


Crawford, E.~J., Filipovi{\'c}, M.~D. and Payne, J.~L.: 2008a, \journal{Serb. Astron. J.}, \vol{176}, 59. 

Crawford, E.~J., Filipovi{\'c}, M.~D., De Horta, A.~Y., Stootman, F.~H., Payne J.~L.: 2008b, \journal{Serb. Astron. J.}, \vol{177}, 61.

Crawford, E.~J., Filipovi{\'c}, M.~D., Haberl, F., Pietsch, W., Payne, J.~L., De Horta, A.~Y.: 2010, \journal{Astron. Astrophys.}, \vol{518}, A35. 

De Horta, A. Y., Filipovi\'c, M. D., Bozzetto, L. M., Maggi, P., Haberl, F., Crawford, E. J., Sasaki, M., Urosevi\'c, D., Pietsch, W., Gruendl, R., Dickel, J., Tothill, N. F. H., Chu, Y.-H., Payne, J. L. and Collier, J. D., 2012, Astronomy \& Astrophysics, 540, A25


{Dickel}, J.~R. and {Milne}, D.~K., 1994, Proceedings of the Astronomical Society of Australia, \vol{11}, 99

{Dickel}, J.~R. and {Milne}, D.~K., 1995, \AJ, \vol{109}, 200


Desai, K.~M., Chu, Y.-H., Gruendl, R.~A., Dluger, W., Katz, M., Wong, T, Chen, C.-H.~R., Looney, L.~W., Hughes, A., Muller, E., Ott, J. \& Pineda, J.~L., 2010, \journal{The Astronomical Journal}, \vol{140}, 584

%

{Edwards}, Z.~I., {Pagnotta}, A. and {Schaefer}, B.~E., 2012, The Astrophysical Journal Letters, \vol{747}, L19

{Filipovic}, M.~D., {Haynes}, R.~F., {White}, G.~L., {Jones}, P.~A., {Klein}, U. and {Wielebinski}, R., 1995, Astron. Astrophys. Suppl. Series, \vol{111}, 311

%
%

Filipovi\'c, M.~D., Pietsch, W., Haynes, R.~F., White, G.~L., Jones, P.~A., Wielebinski, R., Klein, U., Dennerl, K., Kahabka, P., Lazendi{\'c}, J.~S.: 1998, \journal{Astron. Astrophys. Suppl. Series}, \vol{127}, 119.


Gooch, R.: 1995, Astronomical Society of the Pacific Conference Series, \vol{77}, 144

%
%

Grondin, M.-H.; Sasaki, M.; Haberl, F.; Pietsch, W.; Crawford, E. J.; Filipovi\'c, M. D.; Bozzetto, L. M.; Points, S.; Smith, R. C., 2012,  \journal{Astron. Astrophys.}, \vol{539}, 15

Haberl, F., Pietsch, W.: 1999, \journal{Astron. Astrophys. Suppl. Series}. \vol{139}, 277.


{Kosenko}, D., {Helder}, E.~A. and {Vink}, J., 2010, Astronomy and Astrophysics, \vol{519}, A11

%

Macri, L. M., Stanek, K. Z., Bersier, D., Greenhill, L. J. \& Reid, M. J., 2006, \ApJ, \vol{652}, 1133.

Maggi, P., Haberl, F., Bozzetto, L. M., Filipovi{\'c}, M. D., Points, S. D., Chu, Y.-H., Sasaki, M., Pietsch, W., Gruendl, R. A., Dickel, J., Smith, R. C., Sturm, R., Crawford, E. J., De Horta, A. Y., 2012,  \journal{Astron. Astrophys.}, \vol{546}, 109

%
%

Mathewson, D.~S., Ford, V.~L., Dopita, M.~A., Tuohy, I.~R., Long, K.~S., Helfand, D.~J.: 1983, \ApJS, \vol{51}, 345.

Mills, B.~Y., Turtle, A.~J., Little, A.~G., Durdin, J.~M.: 1984, \journal{Aust. J. Phys.}, \vol{37}, 321.

Payne, J.~L., White, G.~L., Filipovi{\'c}, M.~D.: 2008, \MNRAS, \vol{383}, 1175.

%
%
%

{Sault}, R.~J., {Teuben}, P.~J. and {Wright}, M.~C.~H., 1995, Astronomical Society of the Pacific Conference Series, \vol{77}, 433

Sault, R.~J., Wieringa, M.~H.: 1994, \journal{Astron. Astrophys. Suppl. Series}, \vol{108}, 585.

%

{Smith}, R.~C., {Kirshner}, R.~P., {Blair}, W.~P. and {Winkler}, P.~F., 1991, \ApJ, \vol{375}, 652

Tuohy, I.~R., {Dopita}, M.~A., {Mathewson}, D.~S., {Long}, K.~S. and {Helfand}, D.~J., 1982, \ApJ, \vol{261}, 473

{Vukotic}, B., {Arbutina}, B. and {Urosevic}, D., 2007, RMXAA, \vol{43}, 33

\endreferences

}

\end{multicols}

\vfill\eject

{\ }



\naslov{MULTIFREKVENCIONA POSMATRA{NJ}A OSTATAKA SUPERNOVIH U VELIKOM MAGELANOVOM OBLAKU --
SLUQAJ {\bf \SNR} } 


\authors{L.~M. Bozzetto,$^1$ M.~D.~Filipovi\'c,$^1$ D. Uro{\v s}evi{\'c}$^{2}$ \& E.~J.~Crawford$^1$}

\vskip3mm


\address{$^1$University of Western Sydney
 \break Locked Bag 1797, Penrith South DC, NSW 1797, Australia}  
\address{$^2$Department of Astronomy, Faculty of Mathematics, University of Belgrade,
 \break Studentski trg 16, 11000 Belgrade, Serbia}
 \Email{m.filipovic@uws.edu.au}

\vskip3mm


\centerline{\rrm UDK \udc}

\vskip1mm

\centerline{\rit Originalni nauqni rad}

\vskip.7cm

\begin{multicols}{2}

{




\rrm 

Mi predstav{lj}amo nova {\rm ATCA} posmatra{nj}a ostatka supernove (OS) u Velikom Magelanovom Oblaku -- \textrm{\SNR}. Ovaj OC je malih dimenzija ($\sim$8~{\rm pc)} i ima tipiqan spektar sa spektralnim indeksom \mbox{$\alpha=$--0.53$\pm$0.07}, kao i sa vi{\ss}im spektralnim indeksom na severnoj ivici ostatka. 

\textrm{\SNR} emituje nizak nivo radijalno orijentisanog polarizovanog zraqenja na talsnim du{\zz}inama od 3 i 6~cm, karakteristiqnog za mla{dj}e OS. Proce{nj}eno je priliqno jako magnetno po{lj}e u vrednosti od 170~$\mu${\rm G}. Ovaj ostatak je tipiqan rezultat eksplozije supernove tipa {\rm Ia}, dele{\cc}i vi{\ss}e zajedniqkih osobina sa isto tako mladim i u svojim dimenzijama malim OS tipa {\rm Ia}, {\rm LMC SNR J0509-6731}.

}\end{multicols}

\end{document}